\pdfoutput=1

\documentclass[11pt]{article}

\usepackage[
]{acl}

\usepackage{times}
\usepackage{latexsym}

\usepackage[T1]{fontenc}

\usepackage[utf8]{inputenc}

\usepackage{microtype}

\usepackage{inconsolata}

\usepackage{graphicx}

\usepackage{arydshln}
\usepackage{adjustbox}
\usepackage{booktabs}
\usepackage{multicol}
\usepackage{multirow}
\usepackage{amsmath}
\usepackage{amssymb}

\usepackage{times}
\usepackage{siunitx}
\usepackage{latexsym}
\usepackage{tablefootnote}
\usepackage{threeparttable}
\usepackage{pifont}
\usepackage{verbatim}

\usepackage{tcolorbox}
\usepackage{array}
\usepackage{ragged2e}
\usepackage{caption}
\usepackage[T1]{fontenc}
\usepackage[utf8]{inputenc}

\usepackage{url}
\usepackage{multicol}
\usepackage{ulem} 
\usepackage{makecell}
\usepackage{tabularx}
\usepackage{soul}
\usepackage{lscape}
\usepackage{enumitem}
\usepackage{bbm}

\usepackage{float}
\usepackage{nicematrix, tikz}

\usepackage{caption}
\usepackage{subcaption}

\title{Temporal Information Retrieval via Time-Specifier Model Merging}

\author{
 \textbf{SeungYoon Han\textsuperscript{1}}\quad
 \textbf{Taeho Hwang\textsuperscript{1}}\quad
 \textbf{Sukmin Cho\textsuperscript{1}}\quad
 \textbf{Soyeong Jeong\textsuperscript{2}}\quad \\
 \textbf{Hoyun Song\textsuperscript{1}}\quad
 \textbf{Huije Lee\textsuperscript{1}}\quad
 \textbf{Jong C. Park\textsuperscript{1}\thanks{\;Corresponding author}}
 \\
 \textsuperscript{1}School of Computing,
 \textsuperscript{2}Graduate School of AI \\
 Korea Advanced Institute of Science and Technology (KAIST)
\\
\texttt{\{seungyoonee,doubleyyh,nelllpic,starsuzi,hysong,huijelee,jongpark\}@kaist.ac.kr}
}

\begin{document}
\maketitle
\maketitle

\begin{abstract}

The rapid expansion of digital information and knowledge across structured and unstructured sources has heightened the importance of Information Retrieval (IR). While dense retrieval methods have substantially improved semantic matching for general queries, they consistently underperform on queries with explicit temporal constraints--often those containing numerical expressions and time specifiers such as ``in 2015.'' Existing approaches to Temporal Information Retrieval (TIR) improve temporal reasoning but often suffer from catastrophic forgetting, leading to reduced performance on non-temporal queries. To address this, we propose Time-Specifier Model Merging (TSM), a novel method that enhances temporal retrieval while preserving accuracy on non-temporal queries. TSM trains specialized retrievers for individual time specifiers and merges them into a unified model, enabling precise handling of temporal constraints without compromising non-temporal retrieval. Extensive experiments on both temporal and non-temporal datasets demonstrate that TSM significantly improves performance on temporally constrained queries while maintaining strong results on non-temporal queries, consistently outperforming other baseline methods. Our code is available at \url{https://github.com/seungyoonee/TSM}.

\end{abstract}
\section{Introduction}

In the contemporary era of digital information, Information Retrieval (IR)--the process of finding and ranking documents from a large collection that are most relevant to a search query--has become increasingly important as information and knowledge rapidly expand across both structured sources (e.g., knowledge bases) \cite{knowledge_base, temporal_knowledge_base} and unstructured sources (e.g., Wikipedia, web documents) \cite{wikidata}. This significance is more amplified in the era of Large Language Models (LLMs), where IR is a crucial component of Retrieval-Augmented Generation (RAG) \cite{RAG, DBLP:conf/iclr/KhandelwalLJZL20} pipelines.

As the importance of IR continues to grow, there have been significant advances in retrieval methods, notably the development of dense retrieval methods \cite{dpr, contriever}. Dense retrieval leverages neural models to encode both queries and documents into dense embeddings to capture semantic similarity, substantially improving retrieval effectiveness for general-domain queries. However, these models exhibit \emph{attention bias}, where their embeddings are optimized primarily for semantic similarity and topical relevance, making them less effective at capturing temporal expressions in queries \cite{tscontriever}. As a result, dense retrievers struggle with queries containing temporal expressions (e.g., ``in 2015,'' ``between 2010 and 2012'') \cite{timeqa}.


To address these challenges, the field of Temporal Information Retrieval (TIR) has emerged, focusing on improving retrieval accuracy for temporal queries by enhancing temporal understanding capabilities of retrievers \cite{tir1, tir2}. Recent research has attempted to increase the time-awareness of dense models from the pre-training process using different temporal information masking \cite{temporal_masking1, temporal_masking2, temporal_masking3}, fine-tuning process \cite{timeqa, temporal_knowledge_base, tscontriever}. By incorporating temporal awareness, TIR aims to enhance the accuracy and relevance of retrieved documents for temporal queries.

Previous studies have primarily focused on improving retrieval performance for temporal queries, often overlooking the resulting performance drop on non-temporal queries. However, while enhancing temporal retrieval capabilities is important, it is equally crucial to maintain robust performance on non-temporal queries. This is because both temporal and non-temporal queries are fundamentally part of general-domain information retrieval and do not require domain-specific knowledge.

Unlike domain-specific retrieval tasks that target specialized topics, temporal queries remain general in scope, with their distinction based solely on the presence of explicit time constraints--typically signaled by time specifiers such as ``in,'' ``after,'' or ``between.'' Accordingly, this paper treats temporal queries as a subset of general queries with explicit time constraints, while non-temporal queries lack such time specifiers. This distinction highlights the need for retrieval models that can flexibly and effectively handle both query types without sacrificing overall performance.

\begin{figure}
    \centering
    
    \vspace{-0.15in}
    
    \includegraphics[width=\columnwidth]{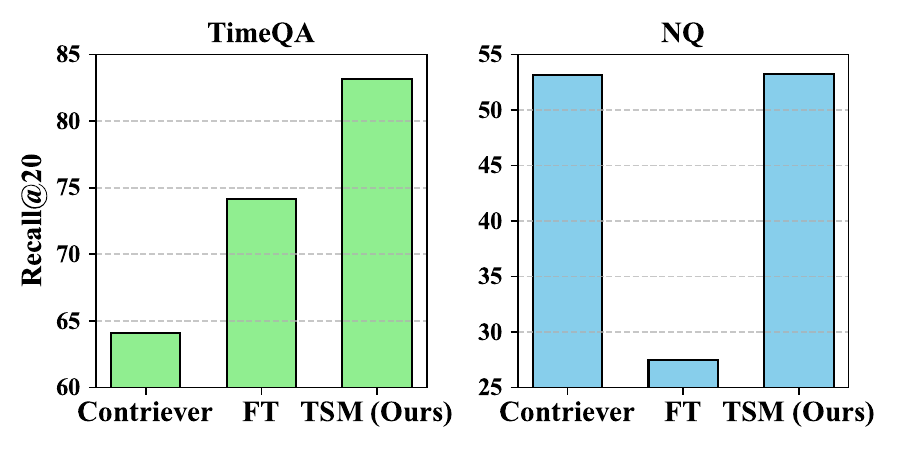}
    
    \vspace{-0.15in}
    
    \caption{Recall@20 performance of vanilla Contriever, full-parameter fine‑tuning (FT), and TSM (Ours) on the temporal dataset TimeQA (green) and the non-temporal general-domain dataset Natural Questions (blue).}
    
    \label{fig:1intro_motivation}

    \vspace{-0.25in}

\end{figure}

Despite this need for balanced retrieval capabilities, fine-tuning dense models to improve accuracy on temporal queries often comes at a significant cost: a noticeable decline in performance on general, non-temporal queries, primarily due to catastrophic forgetting \cite{catastrophic1, catastrophic2}. For instance, as illustrated in Figure \ref{fig:1intro_motivation}, fine-tuning Contriever \cite{contriever} on TimeQA \cite{timeqa} enhances temporal retrieval but substantially reduces performance on the general-domain dataset Natural Questions (NQ) \cite{nq}.

To address this issue, \citet{tscontriever} and \citet{query_router} proposed a routing-based method that directs temporal queries to a temporally fine-tuned retriever and non-temporal queries to a vanilla retriever, which helps mitigate catastrophic forgetting. However, this approach requires maintaining and operating two separate dense retrievers models, resulting in an increased memory usage, which can be resource-intensive in practical deployments. Furthermore, while this method helps preserve performance across both query types, it heavily relies on accurate classification of queries as temporal or non-temporal, which can result in suboptimal retrieval accuracy, as shown in Table \ref{table:5main_result}.



To address the challenge of handling both temporal and non-temporal queries, we propose Time-Specifier Model Merging (TSM), a novel temporal fine-tuning method. TSM involves separately training specialized retrievers on data subsets corresponding to specific time specifiers (e.g., ``in,'' ``after,'' ``between'') for temporal queries with explicit expressions. Each retriever develops expertise in a particular temporal constraint. We then merge these specialized models by simply averaging their parameters, allowing the unified retriever to inherit the specialized performance of each time-specifier-specific model.

This merging process is effective at mitigating catastrophic forgetting because it results in lower-magnitude weight changes--preserving knowledge from both temporal and non-temporal data, rather than overwriting it as in standard fine-tuning \cite{model_merging1, model_merging2}. As a result, the merged model can more effectively encode temporal relevance associated with each time specifier while still maintaining strong performance on non-temporal queries. Extensive experiments on both temporal and non-temporal datasets demonstrate that TSM significantly improves performance on temporal queries while preserving performance on non-temporal datasets. TSM consistently outperforms alternative temporally-aware training methods, including full fine-tuning, regularization, LoRA, routing, and ensembling. 

To summarize, our contributions are threefold:

\begin{itemize}
  \setlength{\itemsep}{5pt}
  \setlength{\parskip}{0pt}
  \setlength{\parsep}{0pt}
\item We identify and address the critical challenge of improving temporal retrieval performance without compromising non-temporal (general-domain) retrieval accuracy, emphasizing the need for retrieval models that can flexibly handle both query types.
\item We propose a novel \textbf{Time-Specifier Model Merging (TSM)} method, which fine-tunes separate, specialized retrievers for individual time specifiers and then merges them into a unified model. This method enables precise handling of temporal constraints while effectively preserving general retrieval capabilities.
\item Through extensive experiments on both temporal and non-temporal datasets, we demonstrate that TSM significantly improves performance on temporal queries without sacrificing non-temporal retrieval accuracy, consistently outperforming other fine-tuning strategies.
\end{itemize}

\section{Related Work}

\noindent{\textbf{Temporal Information Retrieval}} 
Temporal Information Retrieval (TIR) is a specialized subfield of Information Retrieval (IR) focused on accurately interpreting temporal information in both user queries and documents \cite{tir1, tir2}. Temporal information refers to specific points in time (e.g., ``in 2015''), intervals (e.g., ``between 2010 and 2012''), and can be expressed in various forms: \textit{explicit} (e.g., ``January 2010''), \textit{relative} (e.g., ``tomorrow''), or \textit{implicit} (e.g., ``Labor Day'') \cite{tir3}. Temporal queries typically involve time specifiers such as ``after'' or ``between'' to define temporal constraints. TIR research addresses challenges such as temporal query analysis, time-aware embedding, and the extraction of temporal expressions to improve temporal retrieval effectiveness. Our work builds on these developments, aiming to enhance retrieval performance for temporally relevant information, with a focus on \textit{explicit} temporal expressions.

\noindent{\textbf{Semantic vs. Temporal Focus in Dense Models}} Dense retrieval models \cite{dpr, contriever} have advanced Information Retrieval (IR) but still struggle with temporal information retrieval (TIR). This is because their embeddings are primarily optimized for semantic similarity and topical relevance, rather than explicit temporal expressions--a limitation known as \emph{attention bias} \cite{tscontriever}. To address this, recent studies have introduced temporal information masking strategies during pre-training, enabling models to better encode explicit temporal expressions, which leads to improved temporal representations \cite{temporal_masking1, temporal_knowledge_base, temporal_masking2, temporal_masking3}. Other approaches, such as TempRALM, enhance retrievers with temporal scoring mechanisms to more accurately rank documents based on temporal relevance \cite{temp_ralm}. While these methods improve retrieval performance for temporal queries, they often overlook the resulting decline in performance on non-temporal queries.

Among the approaches addressing both temporal and non-temporal retrieval  using off-the-shelf dense models, \citet{tscontriever} and \citet{query_router} proposed a routing-based method that directs temporal queries to a retriever fine-tuned on temporal datasets and non-temporal queries to a vanilla retriever, mitigating catastrophic forgetting. While this preserves performance across query types, it heavily relies on accurate query classification, which can result in suboptimal performance. In this study, we focus on fine-tuning off-the-shelf dense retriever models to handle both temporal and non-temporal queries within a single model, eliminating the dependence on additional modules for query classification.

\noindent{\textbf{Mitigating Catastrophic Forgetting}}
Catastrophic forgetting occurs when a model, after being fine-tuned on a new task or domain, loses performance or knowledge on previously learned tasks \cite{catastrophic1, catastrophic2}. Regularization is a fundamental technique to address this, constraining parameter updates during fine-tuning to preserve pre-trained knowledge \cite{regularization1, regularization2, regularization3}. Low-Rank Adaptation (LoRA) is another effective approach, which introduces a small number of trainable low-rank matrices while keeping most weights frozen \cite{lora}. LoRA and its variants have shown strong performance in continual and out-of-domain learning by isolating task-specific updates and preserving prior knowledge, helping to reduce catastrophic forgetting \cite{lora_ood}.

Another approach is ensemble learning, which combines the predictions of multiple models--each specialized for different tasks or domains -- to achieve balanced performance \cite{ensemble1, ensemble2, ensemble3}. However, this approach requires running multiple models simultaneously, increasing both memory usage and inference costs. Routing-based methods have also been proposed, dynamically directing queries to either a fine-tuned or the vanilla model based on the query type \cite{tscontriever, query_router}. While routing leverages the strengths of both specialized and general models, its effectiveness depends on accurate query classification and still requires maintaining multiple models, making it resource-intensive in practice.

Model merging has recently emerged as a simple and effective approach to mitigating catastrophic forgetting by flattening high-magnitude weight changes during adaptation, resulting in more stable and higher-quality parameter updates \cite{model_merging1, model_merging2}. Motivated by these findings, we adopt model merging in this study and propose a novel temporal fine-tuning method. Our method fine-tunes specialized retrievers for individual time specifiers and merges them into a unified model, enabling effective retrieval for both temporal and non-temporal queries.

\section{Method}

We define the temporal and non-temporal retrieval problem and introduce out method, Time-Specifier Model Merging (TSM). 

\subsection{Problem Formulation and Preliminaries}
We begin by defining the information retrieval task, distinguishing between temporal and non-temporal, and introducing key concepts and notations used throughout our method.

    \noindent{\textbf{Information Retrieval (IR).~}}
    IR identifies a subset of documents $D=\{d_1,d_2,\dots,d_k\}$ from a corpus $C$ that are most relevant to a given user query $q$. Formally, the retrieval process can be defined as:
    \begin{equation}
        D = \{d_1, \dots, d_k\} = \texttt{Retriever}(q, \mathcal{C}),
    \end{equation}    
    where the \texttt{Retriever} function returns the top-$k$ documents from $\mathcal{C}$ ranked by their relevance to $q$.

    \noindent{\textbf{Dense Retrieval.~}}
    Dense retrieval encodes queries and documents into dense vector representations using neural encoders. Let $f_\theta$ denote an encoder parameterized by $\theta$, which maps $q$ and $d_i$ to dense vectors:
    \begin{equation}
        \mathbf{q}=f_\theta(q),\; \mathbf{d}_i=f_\theta(d_i),\:\forall d_i \in \mathcal{C}
    \end{equation}
    The relevance score between a query and a document is computed via the dot product of their vector representations:
    \begin{equation}
        sim(\mathbf{q},\mathbf{d}_i)=\mathbf{q}^\top \mathbf{d}_i
    \end{equation}
    and the retriever selects documents with the highest similarity scores.
    
    \noindent{\textbf{Temporal and Non-Temporal Queries.~}}
    Let $Q$ denote the set of all general-domain queries. The subset of temporal queries $Q_T \subseteq Q$ is defined as:
    \begin{equation}
        Q_T=\{q_T\in Q\:|\:q_T=(s,t),s\in \mathcal{S}, t\in \mathcal{T}\}
    \end{equation}
    where $\mathcal{S}$ is the set of time specifiers: $\mathcal{S}=\{\text{before},\text{between},\dots\}$, and $\mathcal{T}$ is the set of specific temporal point or period: $\mathcal{T}=\{\text{Apr 2020}, \text{[1990,2000]}, \dots\}$. The subset of non-temporal queries $Q_N \subseteq Q$ is given by:
    \begin{equation}
        Q_N=Q\setminus Q_T
    \end{equation}
    such that $Q=Q_T \cup Q_N$ and $Q_T \cap Q_N=\emptyset$.
    
    \noindent{\textbf{Objective of Our Method.~}}
    The objective of our method is to address the newly defined problem of balancing effective temporal retrieval for temporal queries (${Q}_T$) with robust performance on non-temporal queries ($Q_N$), ensuring that improvements in one do not come at the expense of the other. 

\begin{table}[htbp]
    \centering
    
    \vspace{-0.1in}
    
    \begin{adjustbox}{width=\columnwidth}
        \begin{tabular}{lcc}
        
        \toprule
        
        \textbf{Time Specifier} & \textbf{Train} & \textbf{Dev} \\
        
        \midrule
        
        from $[time_1]$ to $[time_2]$       & 11,676 & 2,486 \\
        in $[time]$                         & 5,759  & 1,233 \\
        between $[time_1]$ and $[time_2]$   & 4,888  & 1,054 \\
        after $[time]$                      & 2,741  & 587   \\
        before $[time]$                     & 2,867  & 609   \\
        in early $[time]$s                  & 1,885  & 438   \\
        in late $[time]$s                   & 2,392  & 474   \\
        
        \midrule
        Total                               & 32,208 & 6,881 \\
        
        \bottomrule
        
        \end{tabular}
    
    \end{adjustbox}
    
    \vspace{-0.1in}
    
    \caption{Statistics of the augmented TimeQA dataset showing the number of queries containing each time specifier in the training and development sets.}
    
    \vspace{-0.2in}
    
    \label{table:3method_data_stat}

\end{table}

\subsection{Time-Specifier Model Merging (TSM)}
Now, we introduce our method, TSM, for improving temporal retrieval performance while maintaining strong non-temporal retrieval capabilities. 
TSM first fine-tunes dense retrieval models on data sampled according to each time specifier, and then merges their parameters to create a unified retriever. 

\subsubsection{Data Sampling}\label{data_sampling}
We utilize TimeQA \cite{timeqa} for fine-tuning dense retrievers. Following the TimeQA taxonomy of seven time specifiers--in [$time$], after [$time$], before [$time$], in early [$time$]s, in late [$time$]s, between [$time_1$] and [$time_2$], and from [$time_1$] to [$time_2$]-- we categorize the dataset into seven groups based on these specifiers. Each [$time$] refers to a specific year or a year with a month. However, the original TimeQA training set is imbalanced across the time specifiers. To address this, we use the official TimeQA data processing scrips and annotation labels to augment the comparatively less frequent time specifiers: \textit{after}, \textit{before}, \textit{in early}, and \textit{in late}. As a result, we increase the training set from 25,064 to 32,208 instances and the dev set from 5,348 to 6,881. Detailed statistics for the original dataset are provided in Appendix \ref{appendix:datasets}. Note that we only use answerable questions with gold answers, as non-answerable questions do not have gold answers and therefore cannot be used for contrastive learning, since there would be no positive passages available. Table \ref{table:3method_data_stat} summarizes the statistics of the augmented dataset for each time specifier.


\subsubsection{Specifier-Specific Fine-Tuning}
For each time specifier $s$, we fine-tune a separate dense retriever on the corresponding subset of sampled data. We employ a contrastive learning objective with the InfoNCE loss \cite{contriever}. For a given temporal query $q_T$, the loss is defined as:
\begin{equation*}
\resizebox{\columnwidth}{!}{
    $
    L(q_T,p^+)=-log\frac{e^{sim(q_T,p^+)/\tau}}{e^{sim(q_T,p^+)/\tau}+\sum^n_{i=1} e^{sim(q_T,p^-_i)/\tau}},
    $
}
\end{equation*}
where $p^+$ is the positive passage (containing the gold answer), $\{p^-_i\}^n_{i=1}$ are $n$ in-batch negative \cite{contriever} passages, $sim(q_T,p)$ is the dot-product similarity between the temporal query $q_T$ and passages $p=\{p^+,p^-\}$, and $\tau$ is a temperature hyperparameter that controls the smoothness of the probability distribution.

\subsubsection{Parameter Merging}
After fine-tuning specifier-specific models with parameters ${\theta_1,...,\theta_k}$, we merge them by simply averaging the parameters~\cite{lmcocktail}:
\begin{equation}
    \theta_{merged} = \frac{1}{k}\sum^k_{i=1}\theta_{i}.
\end{equation}
The merged retriever is then used to encode both temporal queries and general, non-temporal queries.

This two-stage approach enables our method to leverage the fine-tuned representations learned from time specifier-specific data while maintaining a merged model for non-temporal retrieval tasks.

\section{Experimental Setups}

\subsection{Datasets}
We evaluate on four QA datasets: two that emphasize \textit{temporal} retrieval--TimeQA \cite{timeqa} and Nobel Prize \cite{tscontriever}--and two representing \textit{non-temporal} retrieval tasks--Natural Questions (NQ) \cite{nq} and MS MARCO \cite{msmarco}. Below, we briefly describe each dataset and clarify our usage protocol.

\textbf{TimeQA} \cite{timeqa} consists of around 25K time-sensitive questions derived from WikiData \cite{wikidata}. These queries focus on facts that evolve over time, requiring models to perform temporal understanding and reasoning. We evaluate on the original TimeQA test set in a closed-domain scenario, using the official document collection chunked by 100-word segments following \citet{chunking} and \citet{dpr}. \textbf{Nobel Prize} \cite{tscontriever} dataset is a template-based corpus created from structured data on Nobel laureates. It includes about 3.2K time-sensitive queries, and we use the provided corpus and test set. \textbf{Natural Questions} \cite{nq} is a benchmark for general QA tasks. We employ the test set from the BEIR benchmark \cite{beir} to evaluate retrieval performance on general queries.
\textbf{MS MARCO} \cite{msmarco} is a widely used benchmark for open-domain question answering. For evaluation, we use its validation set provided through the BEIR benchmark \cite{beir}.


\subsection{Models}
We employ \textbf{Contriever} \cite{contriever} for an \textit{unsupervised} dense retriever, and \textbf{Dense Passage Retriever (DPR)} \cite{dpr} for a \textit{supervised} dense retriever, allowing us to assess the effectiveness of baseline methods and our method on both unsupervised and supervised retrievers.

\begin{table*}[ht!]
    \centering
    \small
    \setlength{\tabcolsep}{5pt}
        \vspace{-0.1in}
    \renewcommand{\arraystretch}{1.1}
    \resizebox{\textwidth}{!}{%
    \begin{tabular}{
        l|cc|cc|cc|cc|cc|cc|cc|cc|cc|cc|cc|cc|cc
    }
        \toprule[1.5pt]
        \multicolumn{1}{c}{\multirow{3}{*}{\textbf{Method}}} &
        \multicolumn{4}{c}{\textbf{TimeQA}} &
        \multicolumn{4}{c}{\textbf{Nobel Prize}} &
        \multicolumn{4}{c}{\textbf{NQ}} &
        \multicolumn{4}{c}{\textbf{MS MARCO}} &
        \multicolumn{4}{c}{\textbf{Average}} \\
        & \multicolumn{2}{c}{\textbf{Recall}} & \multicolumn{2}{c}{\textbf{nDCG}}
        & \multicolumn{2}{c}{\textbf{Recall}} & \multicolumn{2}{c}{\textbf{nDCG}}
        & \multicolumn{2}{c}{\textbf{Recall}} & \multicolumn{2}{c}{\textbf{nDCG}}
        & \multicolumn{2}{c}{\textbf{Recall}} & \multicolumn{2}{c}{\textbf{nDCG}}
        & \multicolumn{2}{c}{\textbf{Recall}} & \multicolumn{2}{c}{\textbf{nDCG}} \\
        & @5 & @20 & @5 & @20
        & @5 & @20 & @5 & @20
        & @5 & @20 & @5 & @20
        & @5 & @20 & @5 & @20
        & @5 & @20 & @5 & @20 \\
        \midrule[1pt]
        \multicolumn{21}{c}{\textit{Unsupervised Dense Retriever}} \\
        \midrule[1pt]
        Contriever
        & 35.29 & 64.07 & 22.98 & 31.49
        & 21.20 & 51.40 & 22.34 & 33.58
        & \underline{29.28} & \underline{53.13} & \underline{21.27} & \underline{28.51}
        & \underline{25.24} & \textbf{45.99} & \underline{17.14} & \textbf{23.20}
        & 27.75 & 53.65 & 20.93 & 29.20 \\
        FT
        & 57.40 & 71.12 & 45.20 & 49.25
        & 14.94 & 39.31 & 14.05 & 23.46
        & 11.32 & 22.69 & 7.75 & 11.10
        & 13.80 & 24.97 & 9.58 & 12.80
        & 24.37 & 39.52 & 19.15 & 24.15 \\
        FT + Reg
        & 60.30 & 74.38 & 46.93 & 51.10
        & 20.21 & 51.21 & 18.67 & 30.65
        & 13.60 & 27.43 & 9.44 & 13.52
        & 15.87 & 28.88 & 10.96 & 14.68
        & 27.50 & 45.48 & 21.50 & 27.49 \\
        LoRA
        & \underline{65.20} & \underline{80.20} & \underline{49.63} & \underline{54.13}
        & 11.04 & 27.52 & 11.54 & 17.52
        & 27.06 & 44.69 & 20.09 & 25.47
        & 20.40 & 37.17 & 14.14 & 18.98
        & 30.93 & 47.40 & 23.85 & 29.03 \\
        Routing
        & 50.15 & 74.35 & 35.36 & 42.54
        & 25.96 & 62.42 & 26.47 & 40.22
        & \underline{29.28} & \underline{53.13} & \underline{21.27} & \underline{28.51}
        & 25.09 & \underline{45.71} & 17.04 & \underline{23.08}
        & 32.62 & \underline{58.90} & 25.04 & 33.59 \\
        Ensembling
        & 63.46	& 77.31	& 48.94	& 53.04
        & \underline{34.39} & \underline{71.47} & \underline{35.13} & \underline{49.12}
        & 25.49	& 45.65	& 18.04	& 24.14
        & 22.36	& 39.97	& 15.39	& 20.51
        & \underline{36.43} & 58.60 & \underline{29.38} & \underline{36.70} \\
        \textbf{TSM (Ours)}
        & \textbf{68.73} & \textbf{83.49} & \textbf{53.45} & \textbf{57.89}
        & \textbf{35.33} & \textbf{75.58} & \textbf{35.73} & \textbf{50.83}
        & \textbf{32.58} & \textbf{53.26} & \textbf{23.66} & \textbf{29.95}
        & \textbf{25.26} & 44.28 & \textbf{17.36} & 22.92
        & \textbf{40.48} & \textbf{64.15} & \textbf{32.55} & \textbf{40.40} \\
        \midrule[1pt]
        \multicolumn{21}{c}{\textit{Supervised Dense Retriever}} \\
        \midrule[1pt]
        DPR
        & 29.98 & 48.08 & 21.08 & 26.39
        & 22.58 & 46.52 & 22.91 & 31.69
        & \textbf{58.20} & \textbf{76.55} & \textbf{46.95} & \textbf{52.67}
        & \underline{21.97} & \underline{35.54} & 15.73 & 19.66
        & 33.18 & 51.67 & 26.67 & 32.60 \\
        FT
        & 52.17 & 66.20 & 41.13 & 45.30
        & 13.75 & 34.92 & 13.32 & 21.35
        & 18.55 & 30.69 & 13.83 & 17.51
        & 6.64 & 12.70  & 4.54 & 6.28
        & 22.78 & 36.13 & 18.21 & 22.61 \\
        FT + Reg
        & 49.03 & 64.39 & 38.34 & 42.83
        & 16.33 & 37.86 & 15.67 & 23.72
        & 17.75 & 31.54 & 12.86 & 17.01
        & 7.72 & 13.99  & 5.41 & 7.23
        & 22.71 & 36.95 & 18.07 & 22.70 \\
        LoRA
        & \underline{65.64} & \underline{78.40} & \underline{51.31} & \underline{55.12}
        & 24.56 & 50.26 & 23.98 & 33.62
        & 47.25 & 62.95 & 38.02 & 42.83
        & 17.87 & 30.97 & 12.85 & 16.62
        & 38.83 & 55.65 & 31.54 & 37.05 \\
        Routing
        & 35.25 & 52.34 & 25.21 & 30.24
        & 19.22 & 42.21 & 19.68 & 28.10
        & \textbf{58.20} & \textbf{76.55} & \textbf{46.95} & \textbf{52.67}
        & \underline{21.97} & 35.53 & \underline{15.74} & \underline{19.67}
        & 33.66 & 51.66 & 26.90 & 32.67 \\
        Ensembling
        & 64.11	& 76.67	& 50.38	& 54.11
        & \underline{30.71} & \underline{58.02} & \textbf{30.48} & \underline{40.80}
        & 43.70	& 60.87	& 34.63	& 39.90
        & 19.91	& 33.61	& 14.05	& 18.01
        & \underline{39.61} & \underline{57.29} & \underline{32.39} & \underline{38.21} \\
        \textbf{TSM (Ours)}
        & \textbf{66.61} & \textbf{79.21} & \textbf{52.53} & \textbf{56.30}
        & \textbf{30.78} & \textbf{60.63} & \underline{30.34} & \textbf{41.72}
        & \underline{48.07} & \underline{66.03} & \underline{38.33} & \underline{43.85}
        & \textbf{23.26} & \textbf{37.80} & \textbf{16.64} & \textbf{20.84}
        & \textbf{42.18} & \textbf{60.92} & \textbf{34.46} & \textbf{40.68} \\
        \bottomrule[1.5pt]
    \end{tabular}
    }
        \vspace{-0.1in}
    \caption{Main results across all datasets and methods, evaluated using Recall and nDCG at top-$\{5,20\}$ documents, with averages reported for each metric. Results are grouped by base retrievers: \textit{Contriever-based} (unsupervised) and \textit{DPR-based} (supervised). The best performance for each metric is shown in \textbf{bold}, and the second-best is \underline{underlined}.}
        \vspace{-0.2in}
    \label{table:5main_result}
\end{table*}

\subsection{Baselines}
We compare our method, TSM, against the following approaches:

\noindent{\textbf{Vanilla Dense Retrievers.~}} Contriever and DPR, using their off-the-shelf checkpoints without any additional fine-tuning.

\noindent{\textbf{Full-Parameter Fine-Tuning (FT).~}} Fine-tuning full parameters of Contriever and DPR on the entire TimeQA training set, without any sampling based on time specifier.

\noindent{\textbf{FT with Regularization.~}} Full-parameter fine-tuning on the entire TimeQA training set with regularization \cite{regularization1}. Specifically, we use a dropout rate of 0.1 and a weight decay of 0.01 during training. Note that all other methods are trained with the same regularization as it is now fundamental in modern model training.

\noindent{\textbf{Low-Rank Adaptation (LoRA).~}} LoRA fine-tuning \cite{lora} of Contriever and DPR on the entire TimeQA training set.

\noindent{\textbf{Routing.~}} A query router that directs temporal queries to the retriever fully fine-tuned on TimeQA and sends general queries to the vanilla retriever, using the router checkpoint provided by \citet{tscontriever}. The router is a two-layer feedforward neural network trained on TimeQA and Natural Questions (NQ) to perform binary classification of queries as either temporal or non-temporal.

\noindent{\textbf{Ensembling.~}} We combine the outputs of multiple dense retrievers, each trained on a different time specifier. Similarity scores from each retriever are first normalized using min-max normalization for a given query. The normalized scores for each candidate passage are then averaged across retrievers to produce an ensemble score, and passages are ranked accordingly \cite{retriever_ensemble}.

Further implementation details are provided in Appendix \ref{appendix:imp_details}.


\subsection{Evaluation Metrics}\label{metrics}
We report our main results evaluating retrieval performance using two standard metrics: \textbf{Recall} and \textbf{nDCG} at top-$\{5,20\}$ documents. Recall measures the proportion of relevant documents successfully retrieved, while nDCG evaluates the quality of ranking by considering both relevance and position.


\begin{figure*}[ht!]
    \centering

    \vspace{-0.1in}
    
    \begin{adjustbox}{width=\textwidth}

        \includegraphics[width=0.5\textwidth]{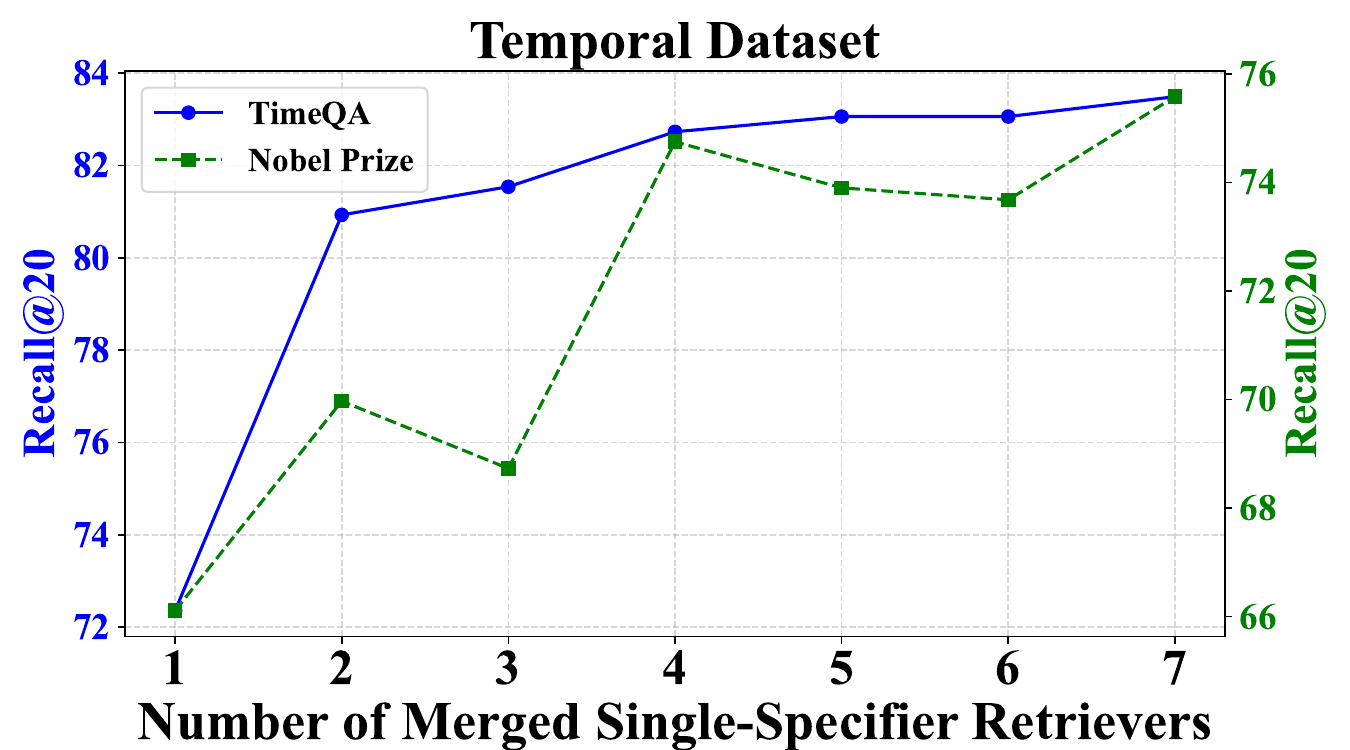}
        \includegraphics[width=0.5\textwidth]{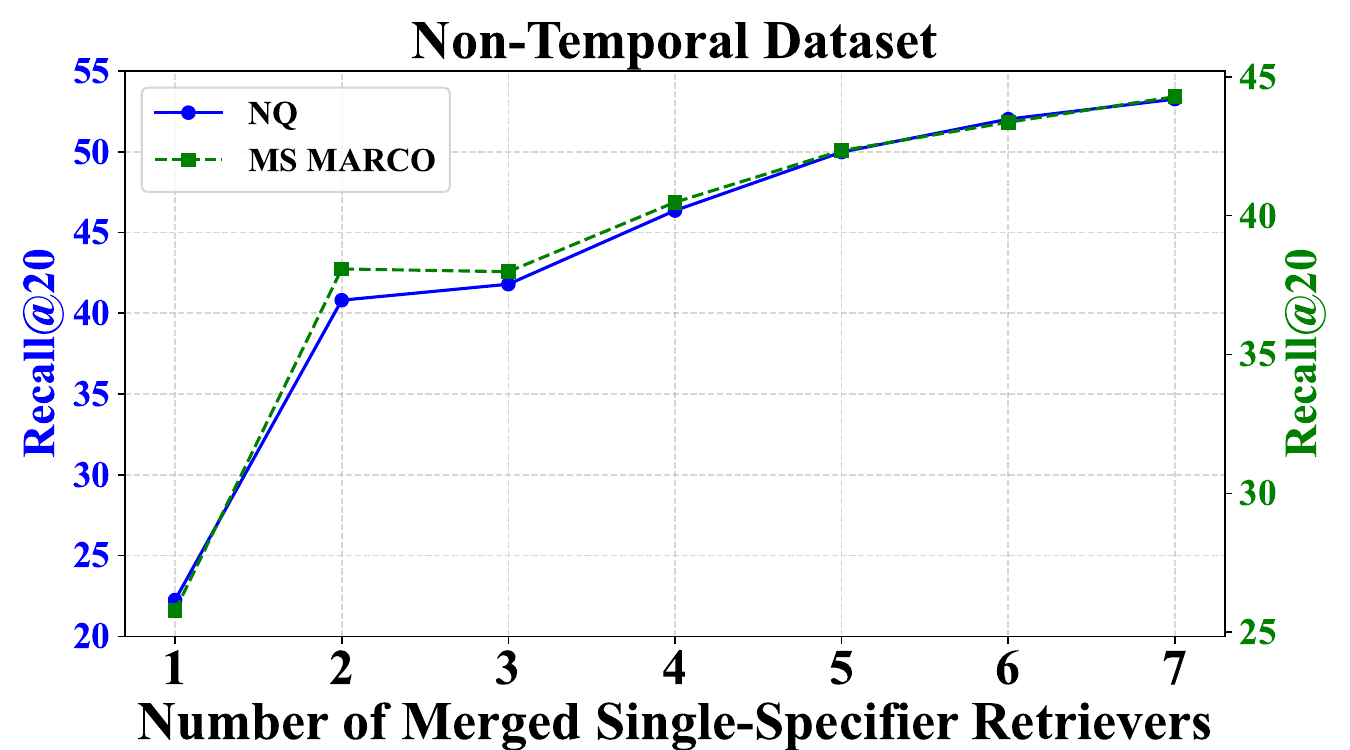}

    \end{adjustbox}

    \vspace{-0.1in}

    \caption{Recall@20 on temporal datasets (TimeQA, Nobel Prize; left) and non-temporal (NQ, MS MARCO; right) datasets as the number of merged single-specifier retrievers increases. 
    }
    

    \label{fig:6-1an_more_merging}
    
\end{figure*}

\definecolor{ForestGreen}{rgb}{0.13,0.55,0.13}
\begin{figure*}[t!]
    \centering

    
    \vspace{-0.1in}

    \begin{adjustbox}{width=\textwidth}
    
        \includegraphics[width=0.5\textwidth]{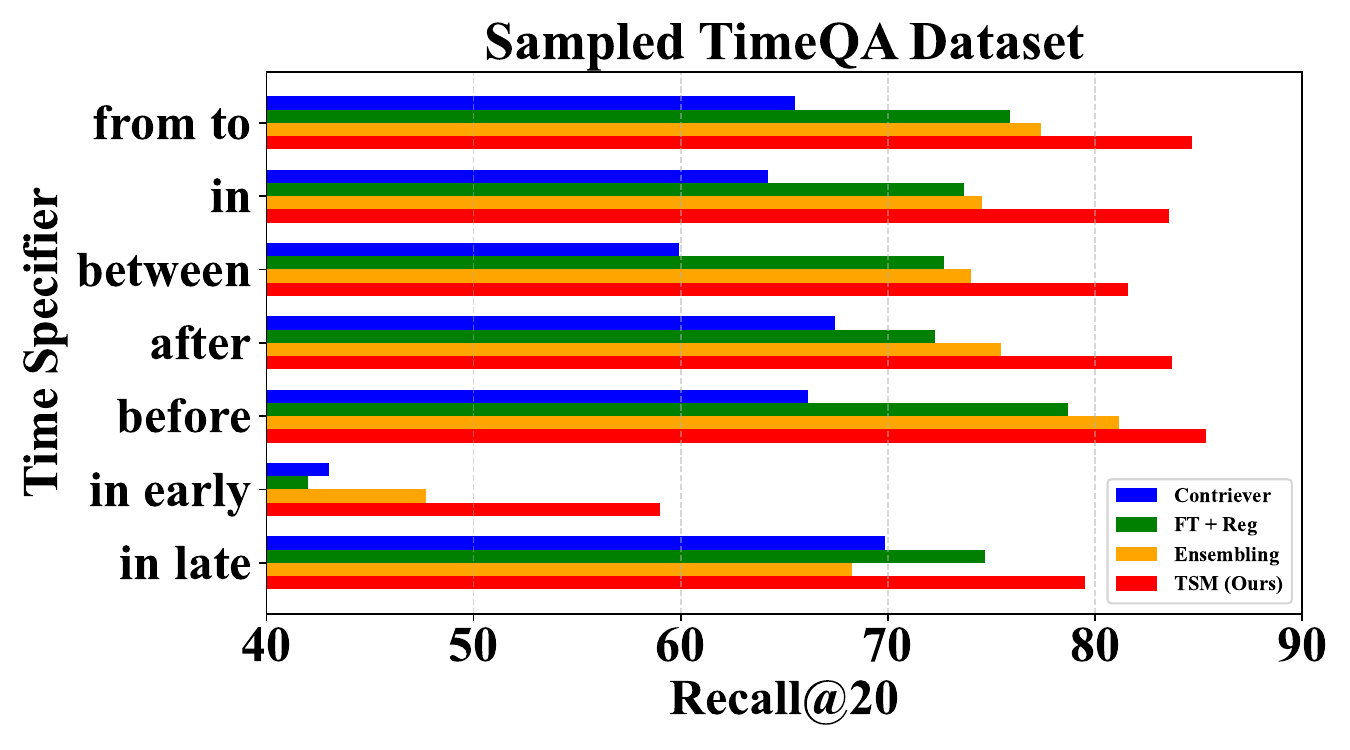}
    
        \includegraphics[width=0.5\textwidth]{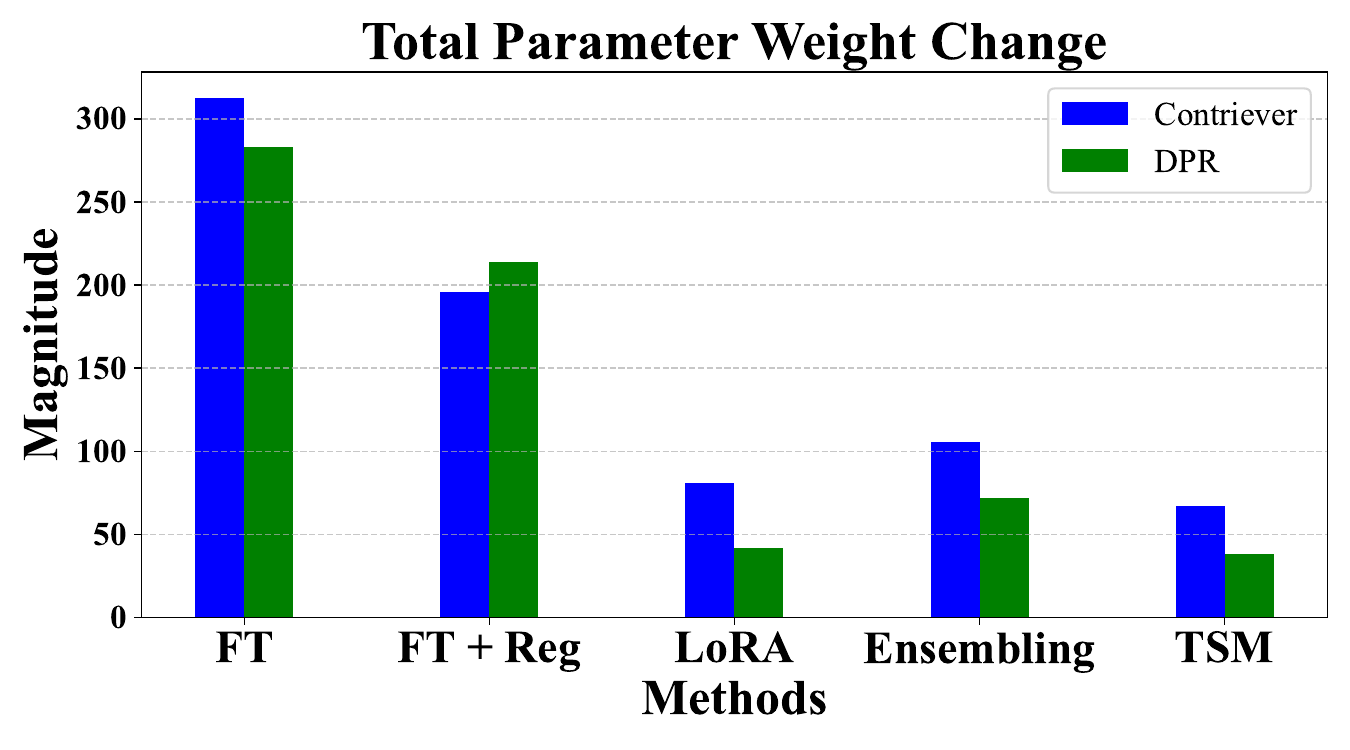}

    \end{adjustbox}
    
    \vspace{-0.1in}
    
    \caption{Left: Recall@20 for each time specifier on the TimeQA test set, comparing vanilla Contriever, FT + Reg, Ensembling, and TSM (Ours). Right: Total parameter weight change after fine-tuning for each method, showing how much all network weights are updated. Lower values indicate more stable parameter adaptation.}
    
    \vspace{-0.2in}

    \label{fig:6an_each_specifier_weight_change}
\end{figure*}
\section{Main Results}

Table \ref{table:5main_result} shows our results across four QA datasets: TimeQA and Nobel Prize as temporal datasets, and NQ and MS MARCO as non-temporal datasets. We evaluate both unsupervised (Contriever) and supervised (DPR) dense retrievers and compare our proposed method, TSM, against several baselines, including vanilla retrievers, full fine-tuning (FT), FT with regularization (FT + Reg), LoRA, routing, and ensembling.

On temporal datasets, TSM achieves the strongest performance across all metrics for both Contriever and DPR. For example, On TimeQA, TSM with Contriever achieves substantial improvements over the vanilla retriever. Similarly, on the Nobel Prize dataset--which serves as an out-of-domain temporal test set--TSM achieves the best performance for both unsupervised and supervised retrievers, confirming its strong generalization to unseen temporal data. Although ensembling yields a marginally higher nDCG@5 on Nobel Prize, TSM remains the most robust performer overall.

On non-temporal datasets, TSM also maintains competitive performance, achieving the strongest results across most metrics with both Contriever and DPR. On NQ, where DPR is trained in-domain, vanilla DPR achieves the highest Recall and nDCG. However, DPR-based TSM performs most closely to vanilla DPR on Recall@5/20 and nDCG@5/20, while outperforming FT, LoRA, and ensembling. On Contriever, which is not trained in-domain, TSM significantly improves retrieval effectiveness. On MS MARCO, which is out-of-domain for both Contriever and DPR, TSM achieves highly competitive performance. For Contriever, it matches or exceeds other baselines on Recall@5 and nDCG@5, and trails slightly behind vanilla and Router on Recall@20 and nDCG@20. Similarly, for DPR, TSM outperforms all other methods across all retrieval metrics. This competitive performance on non-temporal datasets can be attributed to TSM's model merging approach, which reduces the magnitude of weight changes during fine-tuning and helps to preserve non-temporal retrieval capabilities while integrating temporal expertise.

Overall, the average results for both Contriever-based and DPR-based TSM show that TSM consistently outperforms other baselines. These results demonstrate that TSM significantly improves temporal retrieval performance without sacrificing effectiveness on non-temporal queries.

\section{Analyses}

In this section, we systematically examine the effectiveness and underlying mechanisms of our proposed approach.

\subsection{Impact of Merging Specifier-Specific Retrievers}

Figure \ref{fig:6-1an_more_merging} shows how retrieval performance changes as the number of merged single-specifier retrievers increases, for temporal (TimeQA and Nobel Prize) and non-temporal (NQ and MS MARCO) datasets. Single-specifier retrievers are merged sequentially in order of data frequency, from most to least frequent, as shown in Table \ref{table:3method_data_stat}.

For the temporal datasets, Recall@20 improves steadily as more single-specifier retrievers are merged. Specifically, for TimeQA (blue line), Recall@20 starts at approximately 72 with a single retriever and rises to about 83 when all seven retrievers are merged (TSM). The Nobel Prize dataset (green line) shows a similar upward trend, increasing from 66 to 76 as more retrievers are merged.

A comparable trend is observed for the non-temporal datasets. For NQ (blue line), Recall@20 increases consistently from about 22 with one retriever to roughly 53 with all seven merged. MS MARCO (green line) also shows a steady improvement, rising from approximately 26 to 45 as the number of merged retrievers increases.

These results demonstrate that merging multiple retrievers, each trained on a specific time specifier, consistently enhances retrieval performance for both temporal and non-temporal queries.

\newcommand{\hlc}[2][yellow]{{%
    \colorlet{foo}{#1}%
    \sethlcolor{foo}\hl{#2}}%
}



        
        
        
        
        

\begin{table*}[ht!]
    \centering
    
    {\tiny
    
    \begin{tabular}{p{0.1\textwidth} p{0.267\textwidth} p{0.267\textwidth} p{0.267\textwidth}} 
        
        \toprule
        
        & \textbf{Contriever} & \textbf{FT + Reg} & \textbf{TSM (Ours)} \\
        
        \midrule
        
        \textbf{Query} & \multicolumn{3}{p{0.8\textwidth}}{
            Which \hlc[blue!25]{position} did \hlc[blue!25]{Charles Clarke} hold \hlc[green!25]{from May 1997 to May 2001}?
        } \\

        \midrule

        \textbf{Answer} & \multicolumn{3}{p{0.6\textwidth}}{\hlc[yellow!50]{Member of Parliament}} \\
        
        \midrule
        
        \textbf{Top-1\newline Retrieved\newline Passage}
        & \begin{minipage}[t]{\linewidth}
        Guardian Unlimited Politics -- Ask Aristotle: Charles Clarke MP - TheyWorkForYou.com -- \hlc[blue!25]{Charles Clarke} MP - BBC News -- \hlc[blue!25]{Charles Clarke} profile \hlc[green!25]{17 October 2002} - Interview on Meet The Writers, Monocle 24 with Georgina Godwin - \hlc[blue!25]{Charles Clarke} takes a \hlc[blue!25]{leading role in promoting animal protection}. - \hlc[blue!25]{Charles Clarke} interviewed on Blair, Europe and what Gordon Brown must do next. - The Role of Courts in a Democracy: A Debate Video of \hlc[blue!25]{Charles Clarke} in a Public Debate for the Foundation for Law, Justice and Society, Oxford, \hlc[green!25]{2011}
        \end{minipage}
        & \begin{minipage}[t]{\linewidth}
        He was a \hlc[blue!25]{member of the Socialist Campaign Group}, \hlc[blue!25]{Secretary of the All-Party Parliamentary Group} for Vietnam, a \hlc[blue!25]{member of the All-Party Group} on Tibet and \hlc[blue!25]{Chair of the All-Party Parliamentary Group} for Cambodia, \hlc[blue!25]{Member of the Home Affairs Select committee} (\hlc[green!25]{1992--\textbf{97}}), and Chairman of the Home Affairs Select Committee \hlc[green!25]{from \textbf{1997} to 1999} and again \hlc[green!25]{from \textbf{2001} to 2003}.
        \end{minipage} 
        & \begin{minipage}[t]{\linewidth}
        \hlc[blue!25]{Charles} Rodway \hlc[blue!25]{Clarke} (born 21 September 1950) is a \hlc[blue!25]{British Labour Party politician}, who was the \hlc[yellow!50]{Member of Parliament} (MP) for Norwich South \hlc[green!25]{\textbf{from 1997 until 2010}}, and served as \hlc[blue!25]{Home Secretary} \hlc[green!25]{from December 2004 until May 2006}.
        \end{minipage} \\
        
        \midrule
        \textbf{Gold Passage} & No & No & \textbf{Yes} \\
        
        \bottomrule
        
        \end{tabular}
        }
        \vspace{-0.1in}
    
    \caption{Case study comparing retrieved passages using Contriever-based methods: vanilla Contriever, FT + Reg, and TSM (Ours). General, non-temporal information is highlighted in \hlc[blue!25]{blue}, temporal information is highlighted in \hlc[green!25]{green}, and the gold answer that the gold passage should include is highlighted in \hlc[yellow!50]{yellow}. Related information, such as correct temporal information, is in \textbf{bold}.}

    \vspace{-0.2in}
    
    \label{table:case_study}
    
\end{table*}

\subsection{Coverage Analysis Across Specifiers}
Figure \ref{fig:6an_each_specifier_weight_change} (left) compares Contriever, FT + Reg, Ensembling, and TSM on queries grouped by individual time specifiers within the TimeQA test set, reporting Recall@20 for each subset. Across all time specifier categories, TSM achieves the highest recall. For example, on ``between [$time_1$] and [$time_2$]'' queries, TSM outperforms Contriever, FT + Reg, and Ensembling by a significant margin.

Ensembling, which averages the outputs of retrievers fine-tuned on each time specifier, consistently improves performance over single retrievers for every specifier. However, while Ensembling enhances the overall recall, it does not match the level of specialization achieved by model merging. By merging retrievers individually trained on each time specifier, TSM inherits the strengths of each specialist model and more precisely captures the nuances of temporal constraints. This approach avoids the narrow focus of single-specifier retrievers and achieves a more robust temporal understanding than fine-tuning or ensembling.

In summary, while Ensembling provides notable gains by leveraging the diversity of multiple retrievers, model merging (TSM) delivers superior coverage and specialization across all time specifiers, resulting in the best balance between specialization and generality for temporally constrained queries.

\subsection{Parameter Weight Change Magnitude}
Figure \ref{fig:6an_each_specifier_weight_change} (right) shows the total parameter weight change after fine-tuning for each method. Full fine-tuning (FT) and FT with regularization (FT + Reg) result in the biggest weight changes, indicating extensive updates that improve temporal retrieval but also increase the risk of catastrophic forgetting, leading to significant performance drops on non-temporal queries. By contrast, LoRA and Ensembling exhibit much smaller parameter weight changes, reflecting more stable adaptation and a better balance between temporal and non-temporal retrieval. Notably, TSM achieves the smallest parameter changes for both Contriever and DPR, highlighting its effectiveness at integrating temporal expertise while preserving non-temporal retrieval capabilities. The minimal weight change in TSM underscores its ability to mitigate catastrophic forgetting and maintain robust performance across both temporal and non-temporal queries.

\subsection{Case Study: Qualitative Comparison}
Table \ref{table:case_study} presents a case study from the TimeQA test set: ``\textit{Which position did Charles Clarke hold from May 1997 to May 2001?}'' Only TSM successfully retrieved the correct gold passage at top-1, while vanilla Contriever and FT + Reg did not. This qualitative analysis examines the types of information each method prioritizes within the retrieved passages. For clarity, information types are color-coded: temporal features (green), non-temporal features (blue), and the gold answer (yellow).

\textbf{Vanilla Contriever} retrieved a passage with non-temporal information about \textit{Charles Clarke} but lacked explicit temporal details matching the required period. This highlights a tendency to focus on non-temporal content, overlooking crucial temporal context. \textbf{FT + Reg} retrieved a passage containing relevant temporal markers (``\textit{1997}'' and ``\textit{2001}'') but failed to associate them with \textit{Charles Clarke}'s positions, demonstrating a bias toward temporal information at the expense of non-temporal context. \textbf{TSM} retrieved a passage explicitly stating that Charles Clarke was ``\textit{Member of Parliament}'' from \textit{1997} to \textit{2010}, directly addressing both the temporal and non-temporal requirements of the query and fully covering the specified time frame.

This case illustrates three key insights: (1) dense retrievers often overlook temporal information; (2) naïve fine-tuning can shift attention too far toward temporal cues, missing essential context; and (3) TSM's approach of merging time-specifier-specialized retrievers effectively balances temporal and non-temporal information, mitigating attention bias.

\section{Conclusion}

This work addresses the challenge of balancing temporal and non-temporal information retrieval by introducing Time-Specifier Model Merging (TSM), a method designed to address attention bias and catastrophic forgetting. TSM trains specialized retrievers for each time specifier and merges them into a unified model. Experiments on both temporal and non-temporal datasets demonstrate that TSM substantially improves performance on temporally constrained queries while maintaining strong performance on non-temporal queries. Our analysis further show that TSM effectively integrates temporal and non-temporal information, mitigating attention bias and outperforming other baselines. These results establish TSM as a robust and efficient solution for diverse information retrieval tasks.
\section*{Limitations}
While Time-Specifier Model Merging (TSM) demonstrates strong performance in balancing temporal and non-temporal information retrieval, several limitations remain. First, TSM relies on the availability of labeled data for each time specifier; underrepresented or ambiguous temporal expressions may limit the effectiveness of specialized retrievers and the merged model. Second, the current approach focuses on explicit temporal constraints and may not generalize as well to queries with implicit, relative, or underspecified temporal information. Third, our method currently utilizes only seven time specifiers, which may not capture the full range of temporal constraint nuances present in real-world queries. Extending the number and diversity of time specifiers is an important direction for future work to improve coverage and robustness. Fourth, this study merged retrievers solely using simple parameter merging. Alternative approaches leveraging other model merging techniques, such as layer-wise weight averaging \cite{layer-wise_weight_merging} and spherical linear interpolation \cite{slerp} can be further explored. Finally, while our experiments cover several benchmark datasets, further evaluation on more diverse domains and real-world temporal retrieval scenarios is needed to fully assess the generalizability and robustness of TSM.
\section*{Ethics Statement}

This research advances temporal information retrieval by introducing and evaluating the Time-Specifier Model Merging (TSM) method on publicly available benchmark datasets, including TimeQA, Nobel Prize, Natural Questions, and MS MARCO.

We recognize that improved retrieval models, especially those sensitive to temporal constraints, could potentially be misused to surface misleading, outdated, or biased information. To mitigate these risks, we encourage responsible deployment of TSM and recommend incorporating safeguards such as fact-checking and bias detection when applying this technology in real-world systems.

No human subjects, private data, or proprietary information were involved in this research. All model training and evaluation were conducted in accordance with the terms of use of the respective datasets.
\section*{Acknowledgements}

This work was supported by the Institute for Information and communications Technology Promotion (IITP) grant funded by the Korea government (MSIT) (No. 2022-0-00010, Development of Korean sign language translation service technology
for the deaf in medical environment).

\bibliography{acl_latex}

\clearpage

\appendix
\section*{Appendix}

\renewcommand{\thesubsection}{\Alph{subsection}}

\subsection{Additional Experimental Setups}

\subsubsection{Model Weights}

All model weights used for both the vanilla model and training were obtained from Hugging Face as off-the-shelf checkpoints, without any additional training. Below, we provide the exact Hugging Face model names for the weights used in our experiments:

\textbf{Contriever:}

- \texttt{\small{facebook/contriever}}

\textbf{DPR:}

- \texttt{\small{facebook/dpr-question\_encoder-multiset-base}}

- \texttt{\small{facebook/dpr-ctx\_encoder-multiset-base}}

\subsubsection{TimeQA Dataset Statistics}
\label{appendix:datasets}
\begin{table}[ht]
  \centering
  
  \begin{adjustbox}{width=\columnwidth}
  
  \begin{tabular}{lcccc}
    \toprule
    \multirow{2}{*}{\textbf{Time Specifier}}
    & \multicolumn{2}{c}{\textbf{Original}}
    & \multicolumn{2}{c}{\textbf{Augmented}} \\
    
    & \textbf{Train} & \textbf{Dev} & \textbf{Train} & \textbf{Dev} \\
    \midrule
    from $[time_1]$ to $[time_2]$      & 11,676 & 2,486 &     - &    - \\
    in $[time]$                        &  5,759 & 1,233 &     - &    - \\
    between $[time_1]$ and $[time_2]$  &  4,888 & 1,054 &     - &    - \\
    after $[time]$                     &    903 &   201 & 2,741 & 587 \\
    before $[time]$                    &    973 &   181 & 2,867 & 609 \\
    in early $[time]$s                 &    309 &    82 & 1,885 & 438 \\
    in late $[time]$s                  &    473 &    91 & 2,392 & 474 \\
    \midrule
    Total                               & 24,981 & 5,238 & 32,208 & 6,881 \\
    \bottomrule
  \end{tabular}
  
  \end{adjustbox}

  \vspace{-0.1in}
  
  \caption{Statistics for the original and augmented TimeQA datasets illustrate the number of queries containing each time specifier in the training and development sets. To mitigate bias, only the data for the comparatively less frequent time specifiers--after, before, in early, and in late--were augmented.}
  
  \label{table:10appendix_data_stat}

\end{table}

\subsubsection{Temporal Queries in Non-Temporal Datasets}
\begin{table}[ht]

  \centering
  
  \begin{adjustbox}{width=\columnwidth}
  
  \begin{tabular}{lcccc}
  
    \toprule
    
    \multirow{2}{*}{\textbf{Dataset}}
    & \multirow{2}{*}{\textbf{Split}}
    & \textbf{Total}
    & \textbf{Temporal}
    & \textbf{Temporal} \\

    &
    & \textbf{Queries}
    & \textbf{Queries}
    & \textbf{Query (\%)} \\
    
    \midrule
    
    NQ       & Test & 3,452   & 53  & 1.54\% \\
    MS MARCO & Dev  & 509,962 & 232 & 0.05\% \\
    
    \bottomrule
    
  \end{tabular}
  
  \end{adjustbox}

  \vspace{-0.1in}
  
  \caption{Statistics of \emph{explicit} temporal queries within the test splits of two non-temporal datasets, NQ \cite{nq} and MS MARCO \cite{msmarco}. The table reports the total number of queries, the count of temporal queries, and their proportion in each dataset.}
  
  \label{table:11app_non-temporal_dataset}

\end{table}

\subsubsection{Implementation Details}
\label{appendix:imp_details}
For all fine-tuning experiments, each method is trained for five epochs and per-GPU batch size of 64 using on an NVIDIA A100 80GB. We use the publicly available code from \citet{contriever} and follow their hyperparameter settings: a learning rate of 1e-4, the AdamW optimizer \cite{adamw} with a linear learning rate scheduler, and a temperature parameter $\tau$ set to 1.0. Model evaluation is performed every 50 steps based on top-1 accuracy, and the best-performing model is selected accordingly. Additionally, five in-batch negative passages are incorporated in the contrastive learning objective.

\subsection{Additional Experimental Results}

\subsubsection{Coverage Analysis Across Specifiers}
\begin{figure}[ht!]
    \centering
    \includegraphics[width=\columnwidth]{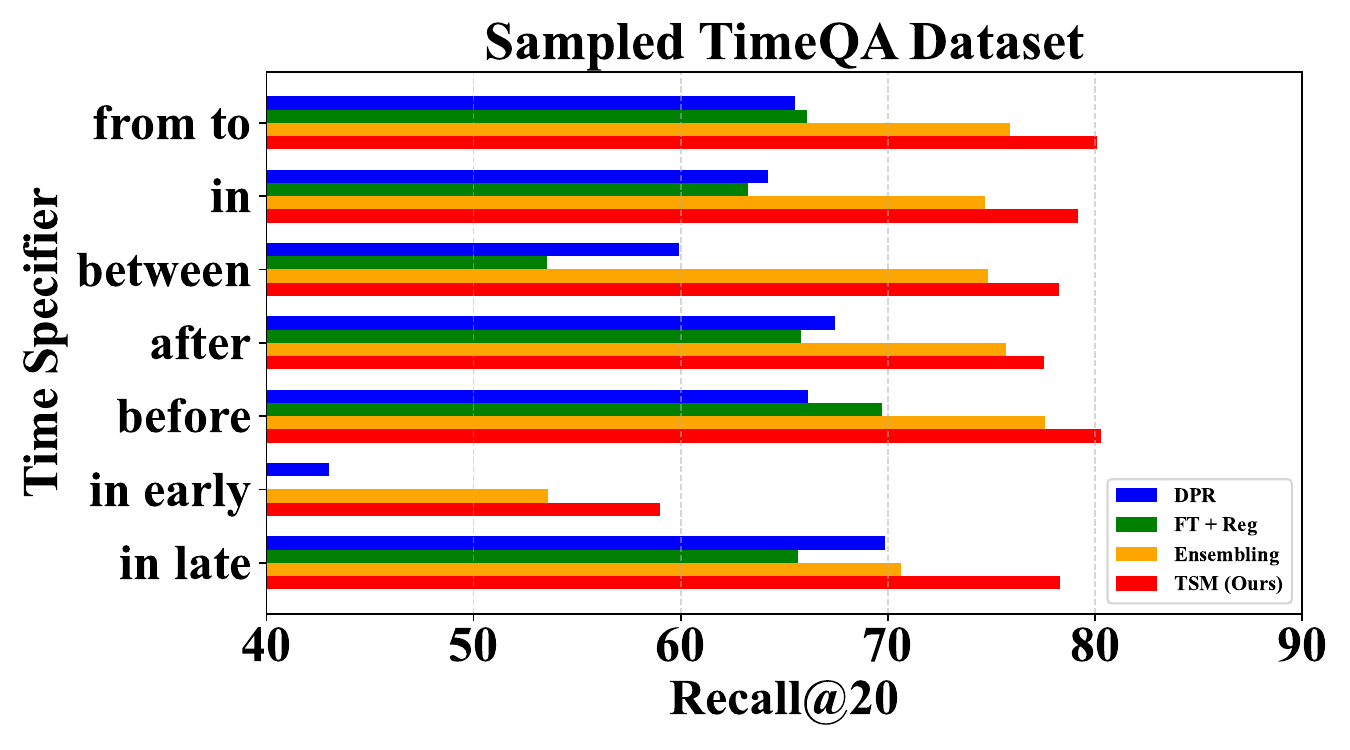}
    \vspace{-0.2in}
    
    \caption{Recall@20 for each time specifier on the TimeQA test set, comparing retrieval performance of vanilla DPR, FT + Reg, Ensembling, and TSM (Ours)}
    
    \label{fig:11app_dpr_each_specifier}
    \vspace{-0.2in}
\end{figure}
Figure \ref{fig:11app_dpr_each_specifier} compares DPR, FT + Reg, Ensembling, and TSM on queries grouped by individual time specifiers within the TimeQA test set, reporting Recall@20 for each subset. Across all time specifier categories, TSM achieves the highest recall. For example, on ``between [$time_1$] and [$time_2$]'' queries, TSM outperforms DPR, FT + Reg, and ensembling by a significant margin.

Ensembling, which averages the outputs of retrievers fine-tuned on each time specifier, consistently improves performance over single retrievers for every specifier. However, while ensembling enhances the overall recall, it does not match the level of specialization achieved by model merging. By merging retrievers individually trained on each time specifier, TSM inherits the strengths of each specialist model and more precisely captures the nuances of temporal constraints. This approach avoids the narrow focus of single-specifier retrievers and achieves a more robust temporal understanding than simply fine-tuning or ensembling.

In summary, while ensembling provides notable gains by leveraging the diversity of multiple retrievers, model merging (TSM) delivers superior coverage and specialization across all time specifiers, resulting in the best balance between specialization and generality for temporally constrained queries.

\definecolor{ForestGreen}{rgb}{0.13,0.55,0.13}
\begin{figure*}[htbp]


    \centering

    \vspace{-0.1in}
    
    \includegraphics[width=\linewidth]{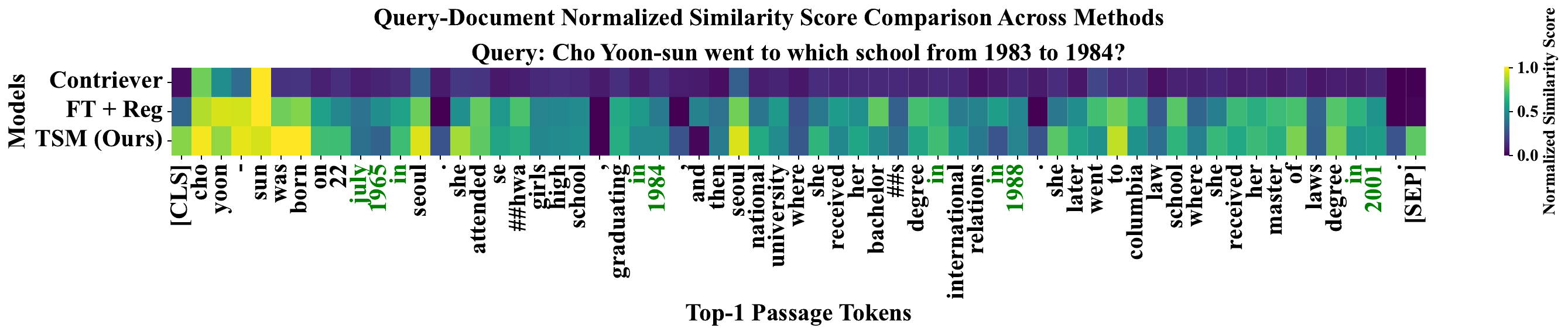}
    
    \vspace{-0.1in}
    
    \caption{Heatmap of normalized query document similarity scores for the query “\emph{Cho Yoon-sun went to which school from 1983 to 1984?}” comparing vanilla Contriever, FT + Reg, and TSM (Ours). Passage tokens in \textcolor{ForestGreen}{green} represent temporal information.}
    
    \vspace{-0.1in}

    \label{fig:11an_sim_score}
    
\end{figure*}

\subsubsection{Parameter Weight Change Magnitude of Each Single-Specifier Model}

\begin{table}[ht]
  \centering
  
  \begin{adjustbox}{width=\columnwidth}
  
  \begin{tabular}{lcccc}
    \toprule
    \textbf{Time Specifier}
    & \textbf{Training Set}
    & \textbf{Weight Change} \\

    & \textbf{Size} & \textbf{Magnitude} \\
    
    \midrule
    from $[time_1]$ to $[time_2]$      & 11,676 & 98.41 \\
    in $[time]$                        &  5,759 & 88.17 \\
    between $[time_1]$ and $[time_2]$  &  4,888 & 98.87 \\
    after $[time]$                     &  2,741 & 55.60 \\
    before $[time]$                    &  2,867 & 55.98 \\
    in early $[time]$s                 &  1,885 & 72.16 \\
    in late $[time]$s                  &  2,392 & 56.61 \\
    \midrule
    Ensembling                         &      - & 75.11 \\
    \midrule
    TSM (Ours)                         &      - & 67.41 \\
    \bottomrule
  \end{tabular}
  
  \end{adjustbox}

  \vspace{-0.1in}
  
  \caption{Parameter weight change magnitude for models fine-tuned on individual time specifiers, compared to Ensembling and TSM. The Ensembling value represents the average weight change magnitude across all single-specifier retrievers. Lower values indicate more stable adaptation.}
  
  \label{table:11app_weight_change_specifier}

\end{table}

\subsubsection{Case Study: Query-Document Similarity Score Analysis}


Figure \ref{fig:11an_sim_score} shows a heatmap of normalized similarity scores between the TimeQA query ``\textit{Cho Yoon-sun went to which school from 1983 to 1984?}'' and the same top-1 retrieved passage, comparing Contriever, FT + Reg, and TSM. The $x$-axis represents the tokenized passage.

\textbf{Vanilla Contriever} mainly highlights non-temporal tokens, such as the person (``\textit{Cho Yoon-sun}'') and location (``\textit{Seoul}''), while largely ignoring temporal tokens such as ``\textit{1984}.'' This indicates that without temporal-specific training, Contriever overlooks time constraints and focuses on general keywords. \textbf{FT + Reg} increases attention to temporal information, especially the correct year ``\textit{1984},'' while still attending to non-temporal tokens, though less effectively than TSM. This demonstrates that temporal fine-tuning helps the model better align temporal aspects of queries and passages. \textbf{TSM} further sharpens this focus, concentrating on the relevant temporal token ``\textit{1984}'' and reducing attention to irrelevant years, while also maintaining strong attention to non-temporal features. This indicates a more balanced integration of temporal and non-temporal information.

Overall, these results show that while Contriever neglects temporal cues, FT + Reg improves temporal sensitivity, and TSM achieves the best balance, accurately attending both temporal spans and key non-temporal details. This balanced attention enables TSM to deliver robust retrieval performance for both temporal and non-temporal queries.

\end{document}